\documentclass{article}
\usepackage{graphicx} 

\usepackage{mathtools}

\DeclarePairedDelimiter\ket{\lvert}{\rangle}
\DeclarePairedDelimiterX\braket[2]{\langle}{\rangle}{#1\,\delimsize\vert\,\mathopen{}#2}

\title{ObfusQate: Unveiling the First Quantum Program Obfuscation Framework}

\usepackage{authblk}
\author[1]{Nikhil Bartake}
\author[2]{See Toh Zi Jie}
\author[3]{Carmen Wong Jiawen}
\author[4]{Michael Kasper}
\author[5]{Vivek Balachandran}
\affil[1,2,3,5]{Singapore Institute of Technology}
\affil[4]{Fraunhofer Singapore}

\usepackage[backend=biber,style=numeric]{biblatex}
\usepackage{xcolor}
\usepackage{soul}
\usepackage{amsmath,amssymb}

\usepackage[margin=1in]{geometry}
\usepackage{caption}
\begin{document}

\maketitle

\begin{sloppypar}

\begin{abstract}
  This paper introduces ObfusQate, a novel tool that conducts obfuscations using quantum primitives to enhance the security of both classical and quantum programs. We have designed and implemented two primary categories of obfuscations: quantum circuit level obfuscation and code level obfuscation, encompassing a total of eight distinct methods. Quantum circuit-level obfuscation leverages on quantum gates and circuits, utilizing strategies such as quantum gate hiding and identity matrices to construct complex, non-intuitive circuits that effectively obscure core functionalities and resist reverse engineering, making the underlying code difficult to interpret. Meanwhile, code-level obfuscation manipulates the logical sequence of program operations through quantum-based opaque predicates, obfuscating execution paths and rendering program behavior more unpredictable and challenging to analyze. Additionally, ObfusQate can be used to obfuscate malicious code segments, making them harder to detect and analyze. These advancements establish a foundational framework for further exploration into the potential and limitations of quantum-based obfuscation techniques, positioning ObfusQate as a valuable tool for future developers to enhance code security in the evolving landscape of software development. To the best of our knowledge, ObfusQate represents the pioneering work in developing an automated framework for implementing obfuscations leveraging quantum primitives. Security evaluations show that obfuscations by ObfusQate maintain code behavior with polynomial overheads in space and time complexities. We have also demonstrated an offensive use case by embedding a keylogger into Shor's algorithm and obfuscating it using ObfusQate. Our results show that current Large language models like GPT 4o, GPT o3 mini and Grok 3 were not able to identify the malicious keylogger after obfuscation. 
\end{abstract}

\section{Introduction}
Code obfuscation using quantum primitives is an underexplored area that holds significant potential for enhancing software security. By incorporating quantum mechanisms into obfuscation methods, we introduce additional layers of complexity that make reverse engineering substantially more challenging. The necessity for adversaries to solve challenges posed by quantum principles such as superposition, entanglement, and quantum gates to reverse engineer the code enhances code security. Currently, there is no quantum obfuscation framework that is capable of obfuscating quantum programs or introducing obfuscations based on quantum principles in classical programs. Our work aims to bridge this gap by developing methods and tools that obscure quantum circuits and program control flows using obfuscations based on quantum primitives, thereby safeguarding sensitive algorithms and data. This approach not only advances secure programming practices, but also fosters the convergence of classical and quantum computing disciplines. This paper discusses the details of the techniques and their implementation in our quantum obfuscation framework, ObfusQate.

\section{Literature Review}
As quantum computing advances, securing quantum algorithms and protecting intellectual property is increasingly critical. Quantum obfuscation techniques offer essential solutions to safeguard circuits from unauthorized access, reverse engineering, and malicious tampering. Unlike classical obfuscation, which deals mainly with bit-level operations, quantum obfuscation leverages unique properties like superposition and entanglement, adding complexity to protection mechanisms. This section explores the current state of quantum obfuscation algorithms in the literature.

One of the initial works in quantum obfuscation came out in 2016 \cite{alagic2016quantum} where the potential applications of quantum black box obfuscators were studied. The work further extends the impossibility of quantum black box obfuscators. Even though perfect obfuscation is impossible in theory, obfuscation has grown as a major cyber-security tool in all forms of computing platforms in the past decades and our hypothesis is that it will be the case with quantum platforms as well. Some of the early works on quantum obfuscations are based on homomorphic encryption. In quantum homomorphic encryption discussed in \cite{dulek2016quantum}, a computationally weak client sends a ciphertext to a quantum server, such that the quantum server performs a quantum computation on the encrypted data, thus producing an encrypted output that the client can decrypt. Another work \cite{zhang2020quantum} on quantum homomorphic encryption-based obfuscation investigates how the existence of a quantum obfuscator can enable the efficient execution of quantum homomorphic functions.

Various quantum obfuscation techniques have been developed to secure quantum circuits and protect intellectual property. Quantum obfuscation using quantum entanglement leverages properties such as the Clauser-Horne-Shimony-Holt (CHSH) game to create quantum opaque predicates, utilizing entangled qubits to conceal parts of the computation and making the classical algorithm's structure difficult to discern \cite{balachandran2021quantum}. Additionally, quantum state obfuscation from classical oracles utilizes quantum teleportation to create an obfuscated state of a quantum function, which can later be restored by an interpreter, thus hiding the function's details while preserving its behavior \cite{bartusek2024quantum}. Indistinguishability obfuscation of null quantum circuits employs techniques like Learning with Errors (LWE) and post-quantum indistinguishability obfuscation to produce secure ``null circuits'' that reject all inputs, making reverse engineering difficult \cite{bartusek_et_al:LIPIcs.ITCS.2022.15}. Furthermore, obfuscation of pseudo-deterministic quantum circuits focuses on transforming code into an unreadable format while preserving functionality, particularly for circuits that produce fixed classical outputs with high probability, highlighting the potential of quantum obfuscation in safeguarding computations and data \cite{bartusek2023obfuscation}.

A different approach of quantum obfuscation is with the assumption that the third party quantum compilers cannot be trusted. Randomized reversible gate-based obfuscation integrates random quantum gates into the original circuit to obscure functionality and protect against reverse engineering \cite{Das2023RandomizedRG} against untrusted compilers. These methods expect the user to have an ineffective quantum compiler that can restore the functionality with the addition of inverse gates. Secure quantum circuit compilation methodologies protect circuits from untrusted compilers by inserting random circuits; factors such as the type, depth, and insertion point of the random circuit are considered to enhance security \cite{das2024secure}. Further techniques include the use of dummy gates and quantum entanglement. A quantum circuit obfuscation method strategically inserts dummy CNOT gates to corrupt functionality before compilation, hindering reverse engineering. Once the dummy gates are removed post-compilation, the original functionality is restored, albeit with a higher overhead \cite{suresh2021short}.

A common trend amongst the current literature in quantum obfuscation is that they are either theoretical or algorithmic designs that could introduce some obfuscations on specific threat models. There is no quantum obfuscation tool that could obfuscate a quantum program or circuit with multiple obfuscation techniques. In this work we have designed and developed multiple quantum obfuscation techniques which have been integrated into a quantum obfuscation framework, that we call ObfusQate. 

\section{ObfusQate}
This section introduces \textit{ObfusQate}, our tool to apply and evaluate various quantum obfuscation techniques. A beta version of the tool is available at \cite{ObfusQate}. The input to ObfusQate can be provided as a quantum circuit in OpenQASM \cite{cross2022openqasm} for circuit-level obfuscations, or as Python code for code-level obfuscations. There are two broad categories of obfuscation that are available in \textit{ObfusQate}.

\begin{itemize} \item \textbf{Quantum Circuit Obfuscations}: These techniques leverage identity gate sequences and mathematically equivalent transformations that are introduced into the quantum circuit. These techniques introduces complexity into the circuit structures while preserving the intended quantum behavior.\item \textbf{Control Flow Obfuscations}: These methods incorporate quantum primitives such as entanglement and superposition into the control flow of classical code. By embedding opaque predicates and other complexity-increasing constructs, they maintain the same output while increasing the complexity of the code structure making it challenging to reverse engineer. 
\end{itemize}

We use the \textit{Quantum Assembly Language (QASM)} simulator in our experimentation, enabling the simulation and validation of quantum algorithms on classical hardware. 

\section{Quantum Circuit Obfuscations}
This section outlines the specific obfuscation techniques for Quantum circuits that were developed and tested, including their implementation and validation using Qiskit and the QASM simulator. The effectiveness, of all the circuit based obfuscation methods, was validated by comparing the outputs of the original and obfuscated circuits. All results confirmed that the obfuscated circuits produced the same output as the original, ensuring the retention of the original output. Please note that, due to memory constraints, these obfuscation methods have been tested only on circuits up to 29 qubits. However, in principle, they can be extended to larger circuits as computational resources permit. It is also important to note that circuit-based obfuscation achieves optimal results when applied to deterministic circuits with guaranteed outputs.

\subsection{Inverse Gates Circuit Obfuscation}

\begin{figure}
  \centering
  \includegraphics[width=1\linewidth]{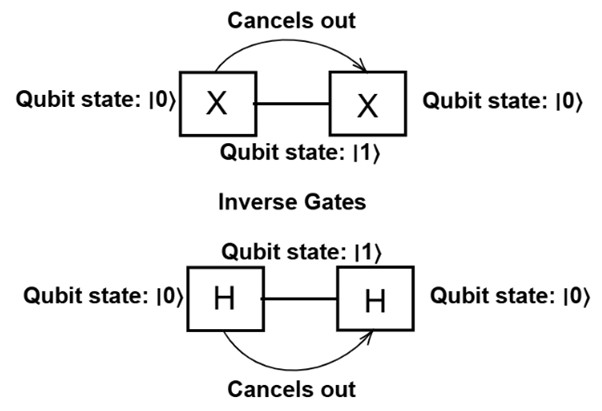}
  \caption{Logic for Inverse Gates}
  \label{fig:inverse_gates}
\end{figure}

The Inverse Gates Circuit Obfuscation technique obscures quantum circuits by inserting pairs of gates and their inverses throughout the circuit. This method is inspired by research on randomized reversible gate-based obfuscation \cite{Das2023RandomizedRG}, which incorporates quantum gates into the original circuit to obscure its functionality and enhance protection against reverse engineering.
 For every gate applied in the circuit, an inverse gate is immediately added thereby nullifying the insertion semantically. This process maintains the original functionality of the circuit but increases its complexity, making reverse engineering more difficult. Figure~\ref{fig:inverse_gates} illustrates this concept with a simple example: applying an `X' gate to a qubit flips its state from ${0}$ to ${1}$, and applying the inverse `X' gate immediately restores it to ${0}$. Similarly, the `H' gate and its inverse maintain the qubit's state. By strategically inserting these gate pairs, the circuit becomes more complex and less understandable to an adversary, while ensuring that the output remains unchanged.

Multiple random gate pairs are applied to the qubits in the circuit. The gates used include single-qubit gates like `H', `X', `Z', and multi-qubit controlled gates like `CX', `CZ', `CCX'. These gate pairs are applied sequentially to ensure that the obfuscation does not alter the logical operations of the circuit but makes it more complex to analyze. The full set of inverse gate combinations used by ObfusQate has been provided in Appendix A.

\subsection{Composite Gates Circuit Obfuscation}

\begin{figure}
  \centering
  \includegraphics[width=1\linewidth]{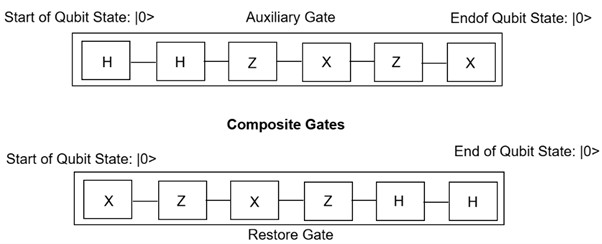}
  \caption{Logic for Composite Gates}
  \label{fig:composite_gates}
\end{figure}

We have developed Composite Gates Circuit Obfuscation technique as a complex version of the Inverse Gate Obfuscation. We group or encompass a set of quantum gates to create a composite gate that we call an Auxiliary gate. The inverse of these gates are encompassed to create another composite gate that we refer to as Restore gate. These are designed in such a way that an Auxiliary gate followed by a Composite gate will result in an identity gate that will restore the quantum states before the insertion of the gates.  The Auxiliary and Restore gates depicted in Figure~\ref{fig:composite_gates} will result in an identity operation. To further complicate the circuit we randomly pick a set of gates from the original circuit and group them together as a composite gate. A reverse engineer will thus have a hard time distinguishing auxiliary gate, restore gate and the extra composite gates containing the original gates.   

\begin{figure}
  \centering
  \includegraphics[width=1\linewidth]{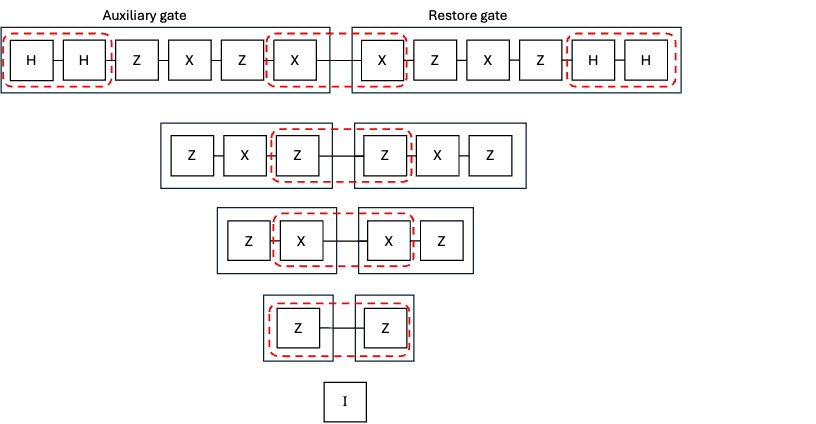}
  \caption{Composite Gates reducing to Identity}
  \label{fig:compositereduction}
\end{figure}

Let us consider the auxiliary sequence $[``H", ``H", ``Z", ``X", ``Z", ``X"]$ and the restore sequence $[``X", ``Z", ``X", ``Z", ``H", ``H"]$. When they are introduced one after another in a quantum circuit it will reduce to an identity. Gates H, X and Z are its own inverse. 
Figure~\ref{fig:compositereduction} shows how these composite gates will reduce to an identity thereby maintaining the original behaviour of the circuit. 

The full set of auxiliary and restore gate combinations used by ObfusQate has been provided in Appendix A.

\subsection{Cloaked Gates Circuit Obfuscation}

\begin{figure}
  \centering
  \includegraphics[width=1\linewidth]{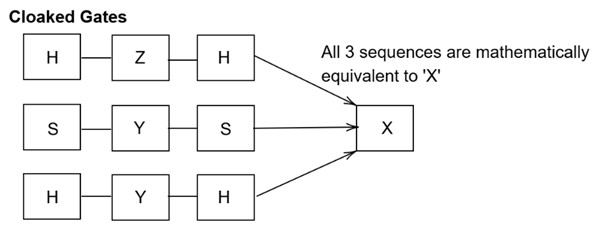}
  \caption{Logic for Cloaked Gates Obfuscation}
  \label{fig:cloaked_gates}
\end{figure}

The Cloaked Gates Circuit Obfuscation technique conceals the genuine functionality of a quantum circuit by replacing original gate operations with dynamically generated substitution sequences. This is an original technique that we developed as these sequences, derived from a predefined substitution map, consist of alternative gate combinations that are mathematically equivalent to the original gates, ensuring the final state of the qubits remains unchanged. Figure~\ref{fig:cloaked_gates} illustrates the logic of this method, using three different substitution maps for the Pauli-X gate.

Each substitution is carefully designed to be equivalent to its original gate. For example, consider one of the substitutions for the Pauli-X gate (\( X \)), (\( HZH \)). We can mathematically demonstrate that they are equivalent. 

The matrix representation for each gate is:
\vspace{-3pt}
\[
H = \frac{1}{\sqrt{2}}
\left[\begin{matrix}
1 & 1 \\
1 & -1
\end{matrix}\right], \quad
Z =
\left[\begin{matrix}
1 & 0 \\
0 & -1
\end{matrix}\right], \quad
X =
\left[\begin{matrix}
0 & 1 \\
1 & 0
\end{matrix}\right]
\]

\[
Z \cdot H = \frac{1}{\sqrt{2}}
\left[\begin{matrix}
1 & 1 \\
-1 & 1
\end{matrix}\right]
\]

\begin{align*}
H (Z H) &= H \cdot (Z H) \\
&= \left( \frac{1}{\sqrt{2}}
\left[\begin{matrix}
1 & 1 \\
1 & -1
\end{matrix}\right] \right)
\cdot \left( \frac{1}{\sqrt{2}}
\left[\begin{matrix}
1 & 1 \\
-1 & 1
\end{matrix}\right] \right) \\
&=
\left[\begin{matrix}
0 & 1 \\
1 & 0
\end{matrix}\right] \\
&= X
\end{align*}

So, \( H Z H \) will give the same quantum rotation provided by the \(X\) gate. By substituting original gates with such mathematically equivalent sequences, the Cloaked Gates method obscures the circuit's functionality without altering its output. This obfuscation method works best for circuits with a high depth of quantum gates.

The full set of cloaked gate combinations used by ObfusQate has been provided in Appendix A.
\subsection{Delayed Gates Circuit Obfuscation}

\begin{figure}
  \centering
  \includegraphics[width=1\linewidth]{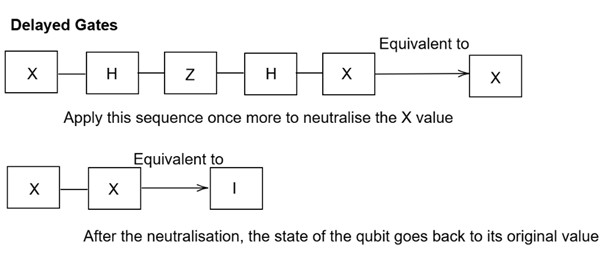}
  \caption{Logic for Delayed Gates}
  \label{fig:delayed_gates}
\end{figure}

The Delayed Gates obfuscation method operates by introducing a set of gates before the original gates to alter the qubit state. These effects are then neutralized by applying corresponding inverse gates after the original set, creating a `delayed' cancellation effect. Since the Qiskit Transpiler has a tendency to decompose simpler gate sequences like inverse gates, this approach was developed as a more complex alternative to resist straightforward optimization. This method was also inspired by \cite{Das2023RandomizedRG}. These gate sequences do not alter the final output of the original circuit but significantly increase the circuit's complexity. By integrating these delayed gates before and after the original gate operations, the method complicates the circuit structure, making it more resistant to reverse engineering and unauthorized analysis. Figure~\ref{fig:delayed_gates} illustrates the logic of this method.

Each identity sequence is mathematically designed to result in the identity operation \( I \) when the gates are reapplied in the specified order after the set
of original gates. Consider the following example sequence that has been used for the Delayed Gates method. This is a sequence that is activated on one qubit:

\textbf{Mathematical Representation:}

Given the sequence: \([``X", ``H", ``Z", ``H", ``X"]\)

1. \textbf{Substitution of \( H \cdot Z \cdot H \) as shown within Cloaked Gates:}
\[
H \cdot Z \cdot H = X
\]

2. \textbf{Substitute this into the sequence}:
\[
X \cdot (H \cdot Z \cdot H) \cdot X = X \cdot X \cdot X
\]
\[
X \cdot X = I, \quad X \cdot X \cdot X = X
\]
The first two \( X \) gates cancel out, resulting in the identity operation. Accounting for the third X gate brings the value back to X.

3. \textbf{Effect of applying the sequence twice:}
\[
(X^3) \cdot (X^3) = X^6 = I
\]

The reversal is achieved by applying a second sequence after the original set of gates. When the original operation commutes with its counterpart copy, the combined effect of the sequences results in (\( I \)). For example, consider a quantum circuit with a single qubit that initially has an \(X\) gate applied. To obfuscate this operation, one might insert an additional gate sequence before the \(X\) gate, such as \([X, H, Z, H, X]\), and then append a corresponding sequence after the \(X\) gate. The sequences are designed such that their overall effect cancels out, i.e. 
\[
[X, H, Z, H, X] \cdot X \cdot [X, H, Z, H, X] = X
\]
The final state of the qubit remains unchanged, while the circuit's structure is significantly obfuscated. There are several delayed gate sequences used in this obfuscation which have been calculated to ensure that the overall transformation remains equivalent to (\( I \)), preserving the original output of the quantum circuit. The full set of delayed gate combinations used by ObfusQate has been provided in Appendix A.

\section{Control Flow Obfuscations}
This section outlines the second type of obfuscation supported by ObfusQate - control flow obfuscation for quantum and classical code.  We have developed a total of four control flow obfuscations based on quantum primitives.

\subsection{Obfuscation via Simple Entanglement}

\begin{figure}
  \centering
  \includegraphics[width=1\linewidth]{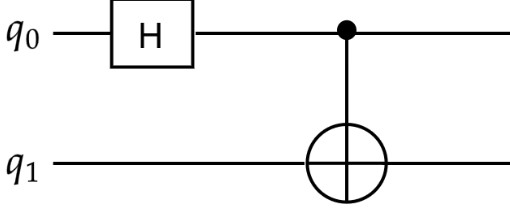}
  \caption{Circuit for Simple Entanglement}
  \label{fig:simple_entanglement}
\end{figure}

Obfuscation via simple entanglement leverages quantum entanglement to obscure the true functionality of a quantum circuit. This method creates an entangled state between qubits, such that the measurement results are correlated and can only yield specific outcomes. By embedding the core logic within branches that are conditionally executed based on the measurement of entangled qubits, the circuit becomes resistant to reverse engineering. Figure~\ref{fig:simple_entanglement} shows a quantum circuit representation of simple entanglement.

To create the entangled state, a Hadamard gate (\( H \)) is applied to qubit $q_{0}$, placing it into a superposition:

\[
H\ket{0} = \dfrac{1}{\sqrt{2}} \left( \ket{0} + \ket{1} \right).
\]

A Controlled-NOT (CNOT) gate is then applied with qubit $q_{0}$ as the control and qubit $q_{1}$ as the target, resulting in the entangled Bell state:
\vspace{-10pt}
\[
|\Phi^+\rangle = \frac{1}{\sqrt{2}} \left( |00\rangle + |11\rangle \right),
\]

Upon measurement, the qubits collapse to either \({00}\) or \({11}\), each with equal probability. This correlation forms the basis of an opaque predicate in the circuit.

In the obfuscation process, the program execution branches are based on the measurement outcomes of the entangled qubits. Both branches contain the same critical functionality, but due to quantum entropy, only one branch executes during a run, and an adversary will not be able to accurately predict which one as there are 2 extra branches correlating to the outcomes, \({01}\) or \({10}\), that will never execute. This inherent uncertainty adds a layer of security, making it challenging to determine the program's true behavior through classical analysis. The unique properties of quantum mechanics used here have no classical equivalent, enhancing the robustness of the obfuscation against reverse engineering.

\subsection{Obfuscation via Entanglement of Variable Qubit Pairs}

\begin{figure}
  \centering
  \includegraphics[width=0.8\linewidth]{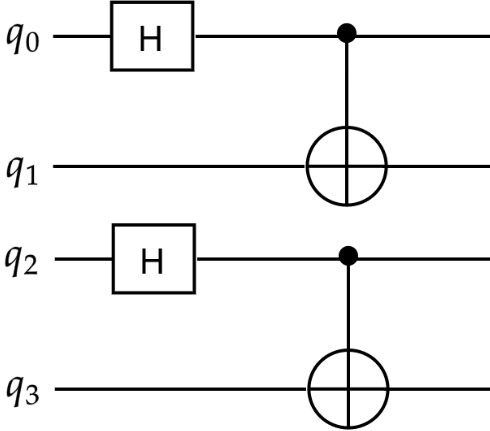}
  \caption{Circuit with Multiple Qubit Pairs} 
  \label{fig:variable_pairs}
\end{figure}

This obfuscation technique builds upon simple entanglement by creating multiple entangled qubit pairs within a quantum circuit. This was inspired by the research within \cite{balachandran2021quantum} which also utilizes entangled qubits to conceal parts of the computation and making the classical algorithm’s structure difficult to discern. This method iterates over a quantum register in steps of two, applying a Hadamard gate (\( H \)) to the first qubit of each pair and a Controlled-NOT (CX) gate to entangle it with the subsequent qubit. By increasing the number of entangled pairs, the probability of measuring all qubits in a specific state decreases exponentially, strengthening the effectiveness of an opaque predicate used for obfuscation. Figure~\ref{fig:variable_pairs} shows a quantum circuit representation for 2 qubit pairs as an example.

As shown in the bell state explanation within simple entanglement, the entangled state has a 50\% probability of measuring $00$ and a 50\% probability of measuring $11$. This equal likelihood can be leveraged in another manner. By creating multiple independent Bell pairs, the probability of measuring all qubits as $\ket{1}$ decreases exponentially. 

\[
P(\text{All qubits are } |1\rangle) = \left( \frac{1}{2} \right)^{n} = \frac{1}{2^{n}}
\]

The value \( n \) is the number of entangled pairs. For example, with eight entangled pairs (\( n = 8 \)), the probability becomes approximately 0.39\%:
\vspace{-2pt}
\[
P = \left( \frac{1}{2} \right)^{8} = \frac{1}{256} \approx 0.39\%.
\]

This low probability allows the opaque predicate to effectively direct program flow into the intended branch with high certainty. The false branch, corresponding to the unlikely event of all qubits being \( |1\rangle \), can contain decoy code or trigger a restart, while the true branch executes the legitimate functionality. By leveraging quantum entanglement and probabilistic measurement outcomes, this obfuscation method introduces inherent unpredictability and complexity. The quantum-based decision mechanism makes it  more difficult to anticipate the program's behavior, thus enhancing security without compromising the original operations.

\subsection{Superposition Shroud Obfuscation}

The Superposition Shroud Obfuscation technique leverages the quantum property of superposition to execute both branches of a conditional statement simultaneously, without collapsing the qubit's state through measurement. A qubit is placed into a superposition state using a Hadamard gate $H$, resulting in a state where the amplitudes of both $\ket{0}$ and $\ket{1}$ are non-zero.

Instead of measuring the qubit, which would force it into one of the basis states, this method accesses the state vector to obtain the amplitudes directly. This allows the program to conditionally execute code based on the presence of both $\ket{0}$ and $\ket{1}$ components in the superposition. 



Mathematically, after applying the Hadamard gate to the initial state $\ket{0}$, the qubit's state becomes:

\[
\ket{\psi} = \frac{1}{\sqrt{2}} \ket{0} + \frac{1}{\sqrt{2}} \ket{1}
\]

The amplitudes \(\frac{1}{\sqrt{2}}\) for both $\ket{0}$ and $\ket{1}$ indicate that the qubit is in an equal superposition of both states. By examining these amplitudes from the statevector, the obfuscation allows both branches of code that depend on $\ket{0}$ and $\ket{1}$ to be executed simultaneously. This technique effectively acts as a reverse opaque predicate, where both branches are true and execute concurrently. It introduces quantum parallelism into the control flow, enhancing security by making it difficult for an adversary to analyze or predict the program's behavior using classical methods.

\subsection{Branch Obfuscation via Superposition}

\begin{figure}
  \centering
  \includegraphics[width=1\linewidth]{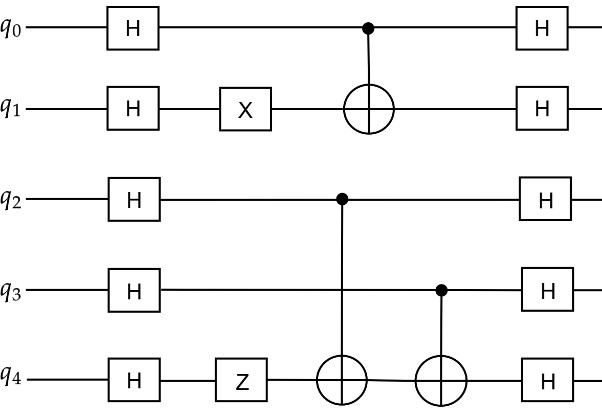}
  \caption{Circuit for Branch Obfuscation via Superposition.} 
  \label{fig:branch_obf}
\end{figure}

This obfuscation technique involves the construction of a quantum circuit where specific qubits are manipulated to produce a deterministic measurement outcome in a manner that obscures the program's control flow.
The process begins by applying Hadamard gates to 5 qubits, all of which begin with an initialization of $\ket{0}$, to place them into superposition, creating a superposed state over all possible basis states. Controlled-NOT (CX) gates are applied to entangle qubits \(q_{2}\) and \(q_{3}\) with qubit \(q_{4}\), an ancillary qubit, thereby establishing correlations between their states. The ancillary qubit facilitates operations on other qubits without being directly measured. Additional gates are used to introduce obfuscations to deter a reverse engineer from calculating the correct output easily. Figure~\ref{fig:branch_obf} shows a quantum circuit representation for branch obfuscation through superposition.

By meticulously orchestrating gate operations, the circuit ensures that measurements yield only specific outcomes—such as qubits \(q_{2}\) and \(q_{3}\) both being 1. This is achieved through quantum interference, where undesired states are canceled out, leaving the desired state with a non-zero probability amplitude. The entanglement created by CX gates, with the ancilla qubit, establishes dependencies that favor the intended measurement outcome. Utilizing superposition and entanglement allows precise control over the probability distribution of results. This deterministic outcome facilitates the implementation of an opaque predicate with multiple program branches, where only one branch evaluates to true during execution while others remain inactive. Developers can embed dead code within these inactive branches without disrupting the main logic. This approach makes it challenging for adversaries to reverse-engineer or predict the program's behavior. By strategically manipulating qubits into specific measurable states, the circuit ensures that only the intended code path is executed.

\begin{figure}
    \centering
    \includegraphics[width=1\linewidth]{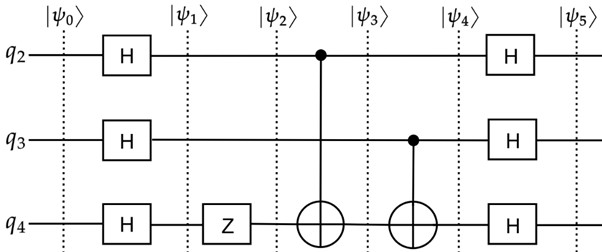}
    \caption{Quantum circuit for qubits $q_{2}$, $q_{3}$ and $q_{4}$}
    \label{fig:Branch Obfuscation}
\end{figure}

Mathematically, the circuit can be split into two segments: one with qubits \(q_0\) and \(q_1\), and the other with the remaining qubits (since no gate connects the qubits from different segments). Consider the segment with qubits \(q_2\), \(q_3\), and \(q_4\) shown in Figure~\ref{fig:Branch Obfuscation}. All qubits are initialized to $\ket{0}$. The following sections clarify the notations used in the equations.

\begin{itemize}
    \item $\{0,1\}^n$ - The set of binary bitstrings of length $n$.
    \item $x \cdot y$ - The bitwise dot product of binary strings $x$ and $y$ of the same length.  
    \item ${CNOT}_{a, b}$ - The CNOT gate applied between qubits $a$ and $b$, with $a$ as the control qubit and $b$ as the target qubit.
    \item $\ket{abc}$ means that $q_{2} = \ket{a}$, $q_{3} = \ket{b}$ and $q_{4} = \ket{c}$.
\end{itemize}

\vspace{-1em}

\vspace{-18pt}
\begin{gather*}
    \begin{aligned}
        \ket{\psi_{0}} &= \ket{000}\\
        \ket{\psi_{1}} &= \big(H \otimes H \otimes H\big) \ket{\psi_{0}} \\&= \frac{1}{2\sqrt{2}}\sum_{x \in \{0,1\}^{3}} \ket{x}\\
        \ket{\psi_{2}} &= \big(I \otimes I \otimes Z\big) \ket{\psi_{1}} \\ &= \frac{1}{2\sqrt{2}} \sum_{x \in \{0,1\}^{3}} {\left(-1\right)^{x \cdot (001)}} \ket{x} \\
        \ket{\psi_{3}} &= \left(\text{CNOT}_{2,4}\right) \ket{\psi_{2}} \\ &= \frac{1}{2\sqrt{2}}\sum_{x \in \{0,1\}^{3}} \left(-1\right)^{x \cdot \left(101\right)} \ket{x} \\
        \ket{\psi_{4}} &= \left(\text{CNOT}_{3,4}\right) \ket{\psi_{3}} \\ &= \frac{1}{2\sqrt{2}}\sum_{x \in \{0,1\}^{3}} \left(-1\right)^{x \cdot \left(111\right)} \ket{x} \\
        \ket{\psi_{5}} &= \left(H \otimes H \otimes H\right)\ket{\psi_{4}} \\ &= \ket{111}
    \end{aligned}
\end{gather*}

Thus, at the end of the circuit, $q_{2}$ and $q_{3}$ are always in the state $\ket{1}$, along with $q_{4}$, which is the ancillary register but the ancillary register's output is not taken into account.

\section{Offensive Use Case: Obfuscating Malicious Payload using ObfusQate}

One of the offensive uses of obfuscation is to obscure malicious code. In this section, we discuss how we embedded a malicious payload into a common quantum algorithm, Shor's factorization, and obfuscate it with ObfusQate. GPT-4o, GPT-o3-mini and Grok-3 were able to identify the malicious payload without obfuscation, but did not succeed in identifying the malicious payload when obfuscated. We demonstrate this technique using a dual-branch architecture: Each branch carries out legitimate quantum computations while subtly embedding malicious behavior, with those hidden operations distributed across both execution paths.

The code stealthily installs a keylogger by splitting its components—writing keystrokes to a file—across two mutually exclusive branches so they never appear together, and then uses an infinite sleep loop to keep itself running indefinitely. We obfuscated this quantum program using the Superposition Shroud method described in Section~5.3.

In the code, the quantum circuit is first initialized and placed in a superposition state by applying the Hadamard gate. The circuit is then executed on a backend and the code executes the two branches:

\begin{itemize}
    \item \textbf{Branch 1:} Branch that defines the classical order‑finding routine and registers a hidden lambda which silently appends data to a file.
    \item \textbf{Branch 2:} Branch that sets up a \texttt{pynput.keyboard.Listener} calling the same lambda on each keystroke—effectively implementing a concealed keylogger.
\end{itemize}

We provided the original code and obfuscated code to ChatGPT-4 and Grok-3  to see if they are able to identify the embedded malicious code in the quantum circuit, with the following prompt.

\textbf{Prompt across all LLM's:} `What does this program do and summarize it in a paragraph'.

Both ChatGPT and Grok succeeded in identifying the malicious payload embedded in the Quantum code when it is not obfuscated. However, both failed to identify the malicious portion when the Quantum code was obfuscated using ObfusQate, as shown in the responses below. This is just one example of how our obfuscation methods can be used to obfuscate potentially malicious code segments.

\begin{itemize}
    \item \textbf{GPT 4o response to original code:} This Python program demonstrates integer factorization using both classical and quantum-inspired methods, with a focus on Shor’s algorithm. It attempts to find a non-trivial factor of a composite integer \( n \) (hardcoded as 381) by either a naive classical order-finding algorithm or a simulated quantum circuit using the \texttt{cirq} library. The quantum method leverages modular exponentiation and quantum phase estimation to compute the order of an integer modulo \( n \), a key step in Shor’s algorithm. If the order \( r \) is even and satisfies certain conditions, it derives a factor from \(\gcd(x^{r/2} \pm 1,\, n)\). Additionally, there's a suspicious component: a keylogger that logs all keypresses to \texttt{C:\textbackslash Temp\textbackslash keystrokes.log}, raising serious security and ethical concerns. The script runs the factorization and then continuously logs keystrokes in the background.

    \item \textbf{GPT 4o response to obfuscated code:}
    This program demonstrates integer factorization using Shor's algorithm, a quantum algorithm that can efficiently find non-trivial factors of a composite number. The script includes both a classical order-finding algorithm (\texttt{naive\_order\_finder}) and a quantum simulation of Shor’s algorithm (\texttt{quantum\_order\_finder}), with the quantum portion using \texttt{Cirq} and \texttt{Qiskit} to simulate quantum circuits, including modular exponentiation and phase estimation. The \texttt{main} function attempts to factor a given integer (in this case, 381) by trying random bases and using the order-finding method to find factors. If a non-trivial factor is found, it prints it; otherwise, it concludes the number is likely prime. The actual quantum circuit is simulated on a classical backend (\texttt{AerSimulator}), which is computationally expensive but serves as a demonstration of how quantum circuits can be used in factoring tasks.

    \item \textbf{GPT-o3-mini response to original code:} This Python script is essentially a demonstration of integer factorization using both a classical (``naive'') order-finding method and a simulated quantum approach (via Cirq) that implements the core of Shor’s algorithm. Given a composite integer \( n \), it repeatedly picks random bases \( x \), computes the order of \( x \) modulo \( n \) (either by brute-force exponentiation or by running a quantum phase estimation circuit), and uses that order to extract a nontrivial factor of \( n \). If \( n \) is even, prime-powered, or trivially shares a factor with a random \( x \), those cases are handled immediately. By default, when run as a script it attempts to factor the hard-coded number 381 using the quantum order finder. Unrelated to its factoring functionality, it also silently starts a background keylogger (using pynput) that appends every keystroke to \texttt{C:\textbackslash Temp\textbackslash keystrokes.log} until the program is terminated.
 
    \item \textbf{GPT-o3-mini response to obfuscated code:}
    This Python program is a demonstration of integer factorization specifically, finding a non-trivial divisor of a composite number-using both a classical (``naive'') method and a simulated quantum order-finding method (the core of Shor’s algorithm). It defines a classical routine (\texttt{naive\_order\_finder}) that iteratively computes the multiplicative order of a randomly chosen base modulo the target integer, as well as a quantum routine (\texttt{quantum\_order\_finder}) built with Cirq (for constructing a modular-exponentiation circuit and performing phase estimation) and Qiskit’s AerSimulator (to simulate the resulting statevector). Both routines feed into a higher-level \texttt{find\_factor} function that implements Shor’s factoring logic: pick random bases, check for trivial factors, determine orders, and compute non-trivial divisors via greatest common divisors. The script’s \texttt{main} function accepts a composite integer (defaulting to 381) and an order-finder choice (naive or quantum) from the command line, then prints any discovered factor before idling until interrupted.

    \item \textbf{Grok 3 response to original code:}
    This Python program implements a factorization demo that attempts to find a non-trivial factor of a given composite integer using either a naive classical algorithm or a quantum circuit based on Shor's algorithm. It uses the \texttt{argparse} library to accept a composite integer \( n \) and an optional order-finding method (`naive' or `quantum') as command-line arguments. The naive method (\texttt{naive\_order\_finder}) computes the order of an integer modulo \( n \) classically, while the quantum method (\texttt{quantum\_order\_finder}) simulates a quantum circuit using the \texttt{cirq} library to perform quantum order finding via Quantum Phase Estimation and modular exponentiation (implemented in the \texttt{ModularExp} class). The program includes helper functions like \texttt{find\_factor\_of\_prime\_power} to handle prime power cases and \texttt{find\_factor} to iteratively attempt factorization by selecting random integers and computing their orders. Additionally, it logs keystrokes to a file using \texttt{pynput}. When run, it outputs whether a non-trivial factor was found or if the number is likely prime, with an example execution factoring 381 using the quantum method. In summary: This program is a factorization tool that takes a composite integer and tries to find a non-trivial factor using either a classical or quantum-inspired approach. It leverages order-finding algorithms, with the quantum version simulating Shor's algorithm using \texttt{cirq}, and includes keystroke logging. It outputs a factor if found or indicates if the input is likely prime, demonstrated with \texttt{n=381} and the quantum method.

    \item \textbf{Grok 3 response to obfuscated code:}
    This program implements a factorization algorithm that can use either a classical or quantum approach to find a non-trivial factor of a given composite integer \( n \). It includes a naive classical order-finding method (\texttt{naive\_order\_finder}) and a quantum order-finding method (\texttt{quantum\_order\_finder}) based on Shor's algorithm, leveraging quantum phase estimation and modular exponentiation. The quantum part simulates a quantum circuit using libraries like Qiskit and Cirq, applying a Hadamard gate to initialize a superposition, followed by modular exponentiation and an inverse quantum Fourier transform to estimate the order \( r \) of a randomly chosen integer \( x \) modulo \( n \). The program then uses this order to attempt factorization by checking if \( x^{r/2} - 1 \) or \( x^{r/2} + 1 \) shares a non-trivial greatest common divisor with \( n \). It includes error handling for invalid inputs, supports command-line arguments to specify the input number and order-finding method, and defaults to factoring 381 using the quantum approach in the provided example. If no factor is found after a set number of attempts (default 30), it suggests the number might be prime. In summary, this Python script demonstrates integer factorization by implementing both a classical and a quantum algorithm (simulated classically), with the latter inspired by Shor's algorithm. It constructs a quantum circuit to find the order of an element modulo \( n \), uses this to identify factors, and provides a flexible framework with command-line options to choose between naive and quantum methods, ultimately attempting to factor the number 381 in the main execution.

\end{itemize}

In conclusion, our work demonstrates that ObfusQate can effectively conceal the control flow within programs. We have shown that models like GPT‑4, GPT-o3-mini and Grok‑3 failed to identify the malicious portion in the obfuscated code.

\section{Evaluation of Obfuscation Overheads}
Obfuscations tend to introduce size and time overheads to a program which is why, in this section we measure these overheads created by the eight obfuscation techniques of ObfusQate. We have selected three quantum algorithms to measure the overheads of ObfusQate: 
\begin{itemize}
    \item Bernstein-Vazirani Algorithm
    \item Shor's Algorithm
    \item Quantum Approximate Optimization Algorithm (QAOA)
\end{itemize}

\subsection{Time overhead of Circuit Obfuscations}
\begin{figure}
    \centering
    \includegraphics[width=1\linewidth]{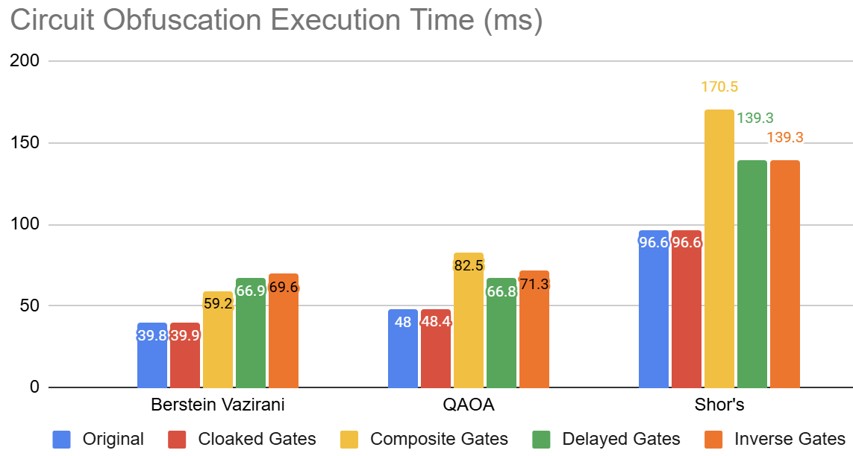}
    \caption{Execution time overhead from circuit-based obfuscation.}
    \label{fig:circuit_time}
\end{figure}

We obfuscated the three algorithms using the four circuit obfuscations provided by ObfusQate. We calculated the time it takes to complete the execution of these circuits in a Quantum simulator and the results are shown in Figure~\ref{fig:circuit_time}. On average, the obfuscations increase the execution time by approximately 43{\%} relative to the original circuits. Notably, the Composite Gates method is the worst overall performer, incurring an average increase of about 66{\%} in execution time across the three algorithms.

\subsection{Time overhead of Code Obfuscations}

\begin{figure}
    \centering
    \includegraphics[width=1\linewidth]{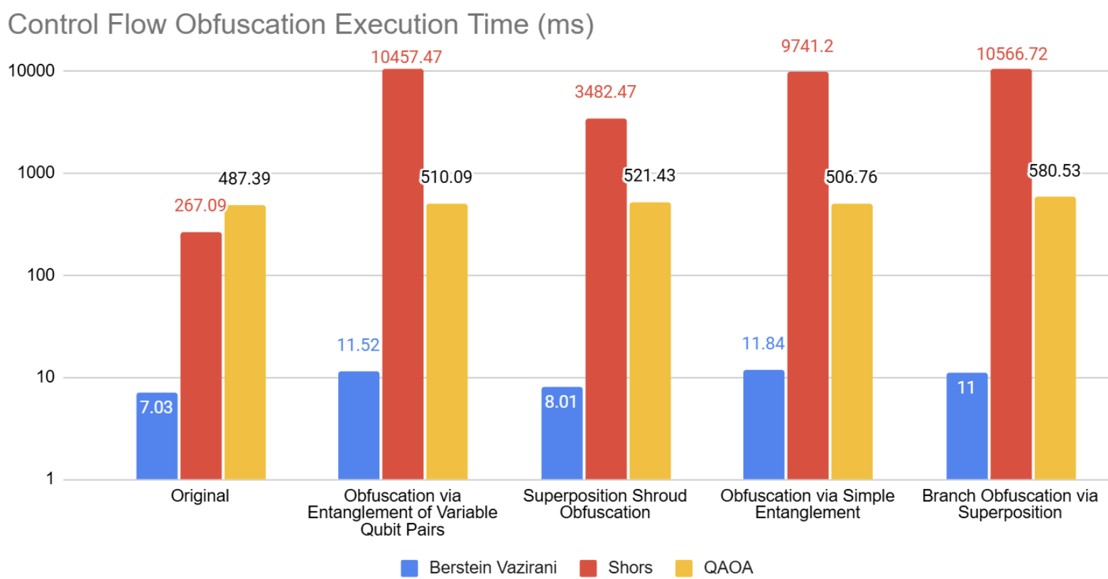}
    \caption{Execution time overhead from control flow obfuscation.}
    \label{fig:controlflow_time}
\end{figure}

For the time overhead incurred by control flow obfuscations you can refer to Figure~\ref{fig:controlflow_time}. On average, the control flow obfuscations increase execution time by roughly 1058{\%} relative to the original programs. The worst-performing method in terms of execution time overhead is Branch Obfuscation via Superposition, which incurs an average increase of about 1323{\%}.
\vspace{-2em}
\subsection{Space overhead of Circuit Obfuscations}

\begin{figure}
    \centering
    \includegraphics[width=1\linewidth]{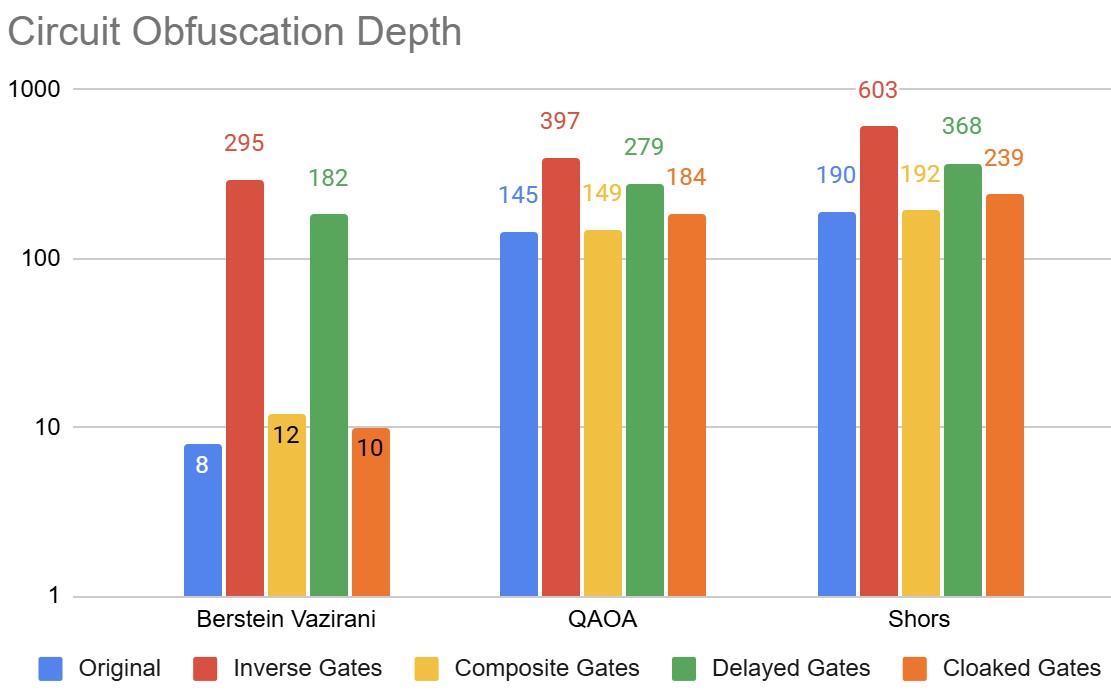}
    \caption{Increase in circuit depth due to circuit-based obfuscation.}
    \label{fig:circuit_depth}
\end{figure}

Figure~\ref{fig:circuit_depth} shows the increase in circuit depth incurred by obfuscations on the three algorithms. We have noticed an average overhead of about 128 additional circuit depth units across all obfuscation methods. Among these methods, Inverse Gates exhibits the highest average increase, adding over 300 circuit depth units in some cases (notably for Shor’s). 

\subsection{Space overhead of Code Obfuscations}

\begin{figure}
    \centering
    \includegraphics[width=1\linewidth]{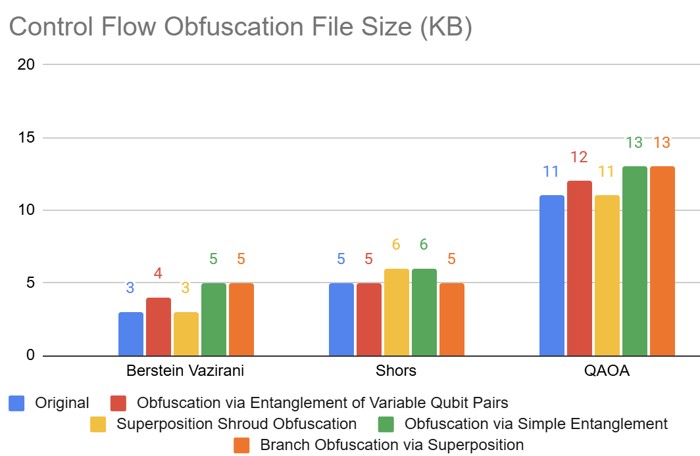}
    \caption{File size increase resulting from control flow obfuscation.}
    \label{fig:controlflow_size}
\end{figure}

Figure~\ref{fig:controlflow_size} shows the file-size overhead introduced by the control flow obfuscations. We observed an average overhead of about 1.5 KB across all obfuscation methods. Among these methods, Branch Obfuscation via Superposition exhibits the highest overall increase, adding up to 7 KB in total when applied to the three algorithms.

\section{ObfusQate Web Interface}
We have developed a centralized web application that provides a user-friendly interface to explore ObfusQate's capabilities \cite{ObfusQate}. The interface includes animations that showcase how the obfuscation functions, which can assist users in understanding and integrating the quantum obfuscation techniques into their code. The GUI supports automatic application of the selected obfuscation and is optimized for accessibility, ensuring consistent performance across various devices.  We hope that this platform will assist developers in seamlessly integrating these obfuscation techniques into their own code bases. Our website artifact is publicly available, and the codes for the obfuscation methods have been uploaded to a GitHub repository at \cite{ObfusQate2024github}.

\section{Future Work}
Our future efforts will focus on further experimental analysis and benchmarking to validate and refine the effectiveness of our obfuscation methods. Through systematic testing across diverse input sets, we aim to ensure that each technique operates as intended without introducing errors or imposing excessive performance overhead. Benchmarking will also enable us to assess metrics such as execution time, memory usage, and scalability, providing a complete view of the operational costs and benefits of each method. Additionally, we plan to implement an obfuscation scheme by deploying custom versions of standard quantum gates. This design increases the structural complexity of the circuit, which can hinder the simplification done by quantum compilers while preserving the intended functionality of the circuit.

\section{Conclusion}
In conclusion, this paper presents eight quantum obfuscation algorithms to enhance code security in quantum computing. Through ObfusQate, we address a critical gap by introducing the first comprehensive quantum obfuscation framework. These advancements not only improve quantum applications but also offer insights for academic and practical efforts in software security and privacy. By implementing these methods, we aim to develop more robust quantum software capable of resisting sophisticated analysis, thereby strengthening security in quantum computing environments.

\section*{Appendix A}
\subsection*{Gate Combinations Used in Inverse Gates}
The following is the full set of gate combinations employed in the inverse gates obfuscation method:
\begin{itemize}
    \item (H, H)
    \item (X, X)
    \item (Z, Z)
    \item (S, Sdg)
    \item (T, Tdg)
    \item (CX, CX)
    \item (CZ, CZ)
    \item (CY, CY)
    \item (CCX, CCX)
\end{itemize}

\subsection*{Composite Gate Sequences}
The following are the gate sequences used in the composite gates obfuscation method:

\paragraph{Auxiliary Gate Sequence}
\begin{itemize}
    \item (H, H, Z, X, Z, X)
\end{itemize}

\paragraph{Restore Gate Sequence}
\begin{itemize}
    \item (X, Z, X, Z, H, H)
\end{itemize}

\subsection*{Cloaked Gates Sequences}
The following sequences are used in the cloaked gates obfuscation method:
\begin{itemize}
    \item (H, Z, H) = X
    \item (S, Y, S) = X
    \item (H, Y, H) = X
    \item (Z, H, Z, H, Z) = X
    \item (SDG, Y, S) = X
    \item (S, Z, Y, Z, S) = X
\end{itemize}

\subsection*{Delayed Gates Sequences}
The following gate sequences are used in the delayed gates obfuscation method:
\begin{itemize}
    \item (Y, S, Y)
    \item (H, S, H, S, H)
    \item (X, H, Z, H, X)
    \item (H, T, T, H, T, T, H)
    \item (Z, H, Y, H, Z)
    \item (S, Z, Sdg)
    \item (T, S, Tdg)
    \item (SWAP, X, SWAP)
    \item (Y, X, Y, X, Y)
\end{itemize}

\end{sloppypar}

\end{document}